\documentclass[12pt,preprint]{aastex}

\shorttitle{The Dynamic Nature of the Adams Ring Arcs}
\shortauthors{Tsui}
\begin{document}
\title{\sc{The Dynamic Nature of the Adams Ring Arcs
\\ - Fraternite, Egalite (2,1), Liberte, Courage}}
\author{K.H. Tsui}
\affil{Instituto de Fisica - Universidade Federal Fluminense
\\Campus da Praia Vermelha, Av. General Milton Tavares de Souza s/n
\\Gragoata, 24.210-346, Niteroi, Rio de Janeiro, Brasil.}
\email{tsui$@$if.uff.br}
\date{}
\pagestyle{myheadings} 
\baselineskip 24pt
\vspace{2.0cm}

\begin{abstract}

By considering the finite mass of Fraternite, the dynamic nature
 of the Adams ring arcs is regarded as caused by the reaction
 of a test body (a minor arc) through the Lindblad resonance (LR).
 Assumming the eccentricity of the test body is larger than that
 of Galatea, this generates several locations along the ring
 in the neighborhood of Fraternite where the time averaged force
 on a test body vanishes. These locations appear to correspond
 to the time dependent configuration of Egalite (2,1), Liberte,
 and Courage, and seem to be able to account for the dynamics
 of the arcs. Such a configuration is a dynamic one because
 the minor arcs are not bounded by the corotation eccentricity
 resonance (CER) externally imposed by Galatea, but are
 self-generated by LR reacting to the external fields.

\end{abstract}

\vspace{2.0cm}
\keywords{Planets: Rings}
\maketitle
\newpage

Since the first observation of the Neptune arcs [Hubbard et al 1986],
 the Voyager 2 mission provided a closed-up measurments of the arcs
 [Smith et al 1989]. Follow-up ground observations have revealled
 changes in arc brightness [Sicardy et al 1999, Dumas et al 1999].
 More recently, these dynamic natures of the arcs are confirmed
 in another ground observation [de Pater et al 2005]. These arcs are
 named Fraternite, Egalite (2,1), Liberte, and Courage. Measuring
 from the center of the main arc Fraternite, they extend a total
 of about $40^{0}$ ahead of Fraternite. According to the currently
 accepted theory, these arcs are confined by the corotation
 resonance potential of the inner moon Galatea because of its
 eccentricity (CER). Orbital parameters are as such that it is
 at the 42/43 resonance giving a resonant site of $8.37^{0}$ on
 the Adams ring [Goldreich et al 1986, Porco 1991,
 Horanyi and Porco 1993, Foryta and Sicardy 1996].
 With Fraternite centered at the potential maximum of CER spaning
 aproximately $5^{0}$ on each side, it appears to fit well the CER
 site. Nevertheless, we remark that the $10^{0}$ span of Fraternite
 contains within it two unstable potential points which ought to
 reduce the angular spread. Furthermore, the minor arcs leading
 ahead of Fraternite and their angular span are mislocated with
 the CER potential maxima.
 In order to account for these minor arcs, the 84/86 corotation
 resonance due to the inclination of Galatea (CIR) is remembered
 giving a potential site of $4.18^{0}$ which offers more options
 in housing the minor arcs. On the other hand, this CIR model
 contradicts directly with the main arc Fraternite.
 While the arc configuration has yet to be resolved in detail,
 recent comparisons among the Voyager and different ground
 observations have shown that the arc intensities are changing
 in time. Occassionally, some arcs flare up and others fade away.
 Furthermore, the arc configuration appears to be changing in time
 as well. The leading arc Courage appears to have leaped over to
 another CER site recently [de Pater et al 2005]. These dynamic
 properties show that the arcs are not in a stable equilibrium
 configuration contrary to the corotation resonance scenario.

Here, we complement the CER model by considering the role of
 Fraternite with its finite mass and eccentricity. The finite mass
 of Fraternite has been suggested by Namouni and Porco [2002] to pull
 on the pericenter precession of Galatea to account for the mismatch
 between the CER pattern speed and the mean motion of the arcs.
 By following on this suggestion, the Lindblad resonance (LR) of a
 test body (a minor arc) under the presence of Fraternite has been
 evaluated. Through the equations of motion of the test body,
 it is shown that the LR of the test body generates locations on
 the Adams ring where the time averaged force acting on it vanishes.
 These locations appear to be compatible to Egalite (2,1) [Tsui 2007].
 In this model, where the direct action of Fraternite surpasses the
 CER potential of Galatea, the arc locations are determined by the
 LR reaction of the arc. For this reason, the arc configuration
 does not have to be static. We extend this same model to include
 also Liberte and Courage for the entire arc system.
 
According to this LR reaction model, only Fraternite is confined
 by the externally imposed CER of Galatea, while the minor arcs
 are hosted at these locations by Fraternite. The locations of the
 minor arcs are given by the roots of Eq.(9) of Tsui [2007] which is
\\
$$4\,\tan(\frac{1}{2}\Delta\theta_{sf})
 \sin(\frac{1}{2}\Delta\theta_{sf})
 \cos[(n+1)\Delta\theta_{sf}-\Delta\phi]\,
 =\,-0.5490\times 10^{8}\frac{m_{f}}{M}\,\,\,,\eqno(1)$$
\\
where $\Delta\theta_{sf}=(\theta_{s}-\theta_{f})$ is the difference
 of the longitudes, $\Delta\phi=(\phi_{s}-\phi_{f})$ is the difference
 of the arguments of perihelion, and $n=42$. The subscripts $s$ and $f$
 denote the test body and Fraternite, and $M$ is the mass of the central
 body Neptune. The third factor, $(n+1)\Delta\theta_{sf}$, on the left
 side is a fast oscillating term that gives $(n+1)$ CER sites along
 the Adams ring. The first two factors, $\Delta\theta_{sf}/2$, on the
 left side are slow oscillating terms that modulate the third factor.
 The left side of Eq.(1), with $\Delta\phi=0$, is plotted in Fig.1
 in thick line. It shows the fast oscillations of the third factor.
 These oscillations grow in amplitude because of the slow modulations
 of the first two factors.
 With mass ratio $m_{f}/M=6.4\times 10^{-10}$, which corresponds to
 $m_{f}=6.4\times 10^{16}$ Kg for Fraternite, the right side of Eq.(1)
 is also indicated in Fig.1 through a straight horizontal line.
 The intercepts of these two plots give the roots of Eq.(1) at
 ($11.8^{0}$ (Egalite 2), $13.8^{0}$ (Egalite 1)),
 ($19.3^{0}$, $22.7^{0}$ (Liberte)),
 ($27.4^{0}$, $31.2^{0}$ (Courage 1999)),
 ($35.7^{0}$, $39.7^{0}$ (Courage 2003)),
 ($44.0^{0}$, $48.1^{0}$), where the corresponding arcs are indicated
 at the estimated locations. The intercepts are grouped in pairs
 within brackets. Each pair comes from a downward cycle of the fast
 oscillating third factor. We have also superimposed a set of constant
 amplitude CER sites in Fig.1, with Fraternite centered at a potential
 maximum, in thin line for comparison and discussion purposes.
 Since the right side of Eq.(1) is much less than unity, the intercepts
 are close to the $y=0$ axis. For small $\Delta\theta_{sf}$, the first
 two factors are most important in determining the nearest intercepts
 corresponding to Egalite (2,1). Nevertheless, the position of these
 two intercepts for Egalite (2,1) are sensitive to the mass variation
 of Fratenite due to the mass factor $(m_{f}/M)$ on the right side.
 They could even disappear should Fraternite be fifty percent more massive.
 They are also sensitive to $\Delta\phi$ althought not so much as the
 mass ratio.
 
As $\Delta\theta_{sf}$ increases, the intercepts are approximately
 given by the third factor
\\
$$\cos[(n+1)\Delta\theta_{sf}-\Delta\phi]\,\simeq\,0\,\,\,,\eqno(2)$$
\\
which are near the mid-points of CER sites, not near the maxima.
 The mid-points are separated by $4.19^{0}$ which reminds us the
 separation of CIRs. But, with Fraternite centered at a CIR maximum,
 the minor arcs would be positioned near the minima of CIRs,
 instead of maxima.
 Although the intercepts are there along the entire Adams ring,
 the action of Fraternite's field gets progressively attenuated
 as such that the arcs can only be confined in its neighborhood.
 This happens to agree with the arc signals that attenuate away
 from Fraternite. The minor arcs indeed get less and less intense
 as they get farther and farther away from Fraternite. 
 Let us now consider the slow change of $\Delta\phi$. This would make
 the cosine function of Eq.(2) drift by half a cycle, and cause the
 pairs of intercepts to drift out by approximately $2^{0}$ which
 could account for the slight variations of the arc positions among
 measurements of different years.
 
Since the roots of Eq.(1) are locations where the force on a test
 body vanishes only on a time averaged base due to its reaction
 to the external fields through LR, this scenario is not one of a
 static equilibrium imposed externally. Test bodies could migrate
 on a long time scale from one site to another leading to
 flaring up of some arcs and fading away of others. This could also
 displace (resonant jump) Courage from $31.2^{0}$ to $39.7^{0}$
 [de Pater et al 2005]. Although there are only arcs in the leading
 positions ahead, it seems to the author that the measurements
 of de Pater et al [2005] tend to indicate also weak signals, under
 large background noise, of arcs in the trailing positions behind.
 It would be consistent to this dynamic model should weak new arcs
 are confirmed in the conjugate positions of Egalite on the trailing
 side.

\clearpage
\begin{figure}
\plotone{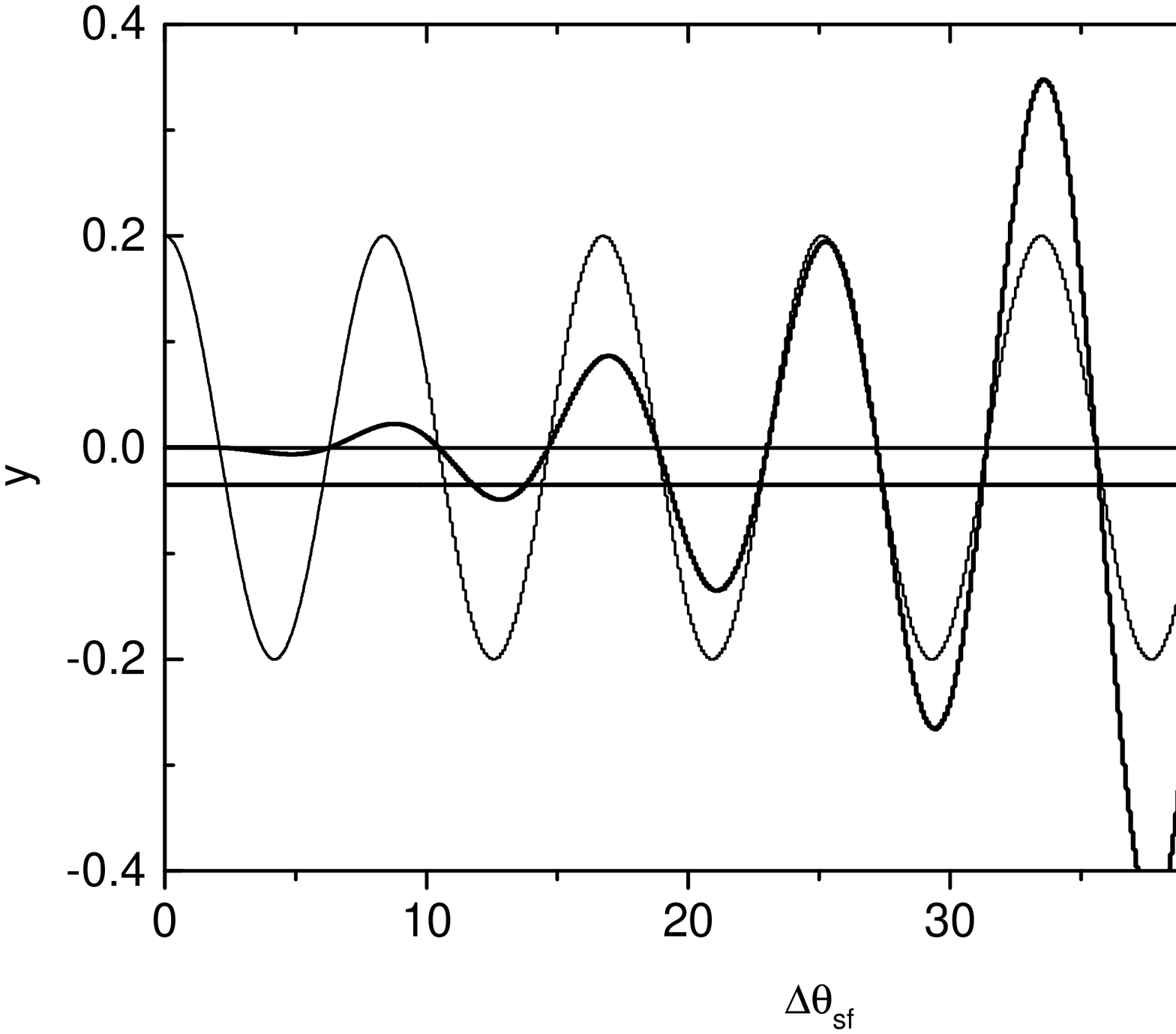}
\caption{The left and the right sides of Eq.(1),
 denoted by the y label, are plotted as a function of
 $\Delta\theta_{sf}$ in degree. The intercepts give the
 roots of Eq.(1) that define the locations where the
 time averaged force on a test body vanishes.
 Constant amplitude CER sites with Fraternite centered
 at a potential maximum are also shown in thin line
 for comparisons.}
\label{fig.1}
\end{figure}

\end{document}